\documentclass[sigconf,screen]{acmart}
\setcopyright{acmcopyright}
\acmPrice{15.00}
\acmDOI{10.1145/3468264.3468553}
\acmYear{2021}
\copyrightyear{2021}
\acmSubmissionID{fse21main-p141-p}
\acmISBN{978-1-4503-8562-6/21/08}
\acmConference[ESEC/FSE '21]{Proceedings of the 29th ACM Joint European Software Engineering Conference and Symposium on the Foundations of Software Engineering}{August 23--28, 2021}{Athens, Greece}
\acmBooktitle{Proceedings of the 29th ACM Joint European Software Engineering Conference and Symposium on the Foundations of Software Engineering (ESEC/FSE '21), August 23--28, 2021, Athens, Greece}

\def\BibTeX{{\rm B\kern-.05em{\sc i\kern-.025em b}\kern-.08emT\kern-.1667em\lower.7ex\hbox{E}\kern-.125emX}}

\usepackage{graphicx}
\usepackage{array}
\usepackage{url}
\usepackage{fancybox}
\usepackage{multirow}
\usepackage{color}
\usepackage{colortbl}
\usepackage{balance}
\usepackage{fancyhdr}
\usepackage{subfig}
\usepackage{diagbox}
\usepackage{bbding}
\usepackage{placeins}
\usepackage{booktabs}
\usepackage{enumitem}
\usepackage{xcolor,lipsum}

\usepackage{textcomp}
\usepackage{algorithmic}
\usepackage{listings}
\usepackage{color}
\usepackage{commath}
\usepackage{comment}

\definecolor{light-gray}{rgb}{.906,  .902,  .902}

\usepackage{bm}

\begin{document}
\title{Automating the Removal of Obsolete TODO Comments}

\begin{abstract}
TODO comments are very widely used by software developers to describe their pending tasks during  software development.
However, after performing the task developers sometimes neglect or simply forget to remove the TODO comment, resulting in obsolete TODO comments. 
These obsolete TODO comments can confuse development teams and may cause the introduction of bugs in the future, decreasing the software's quality and maintainability.
Manually identifying obsolete TODO comments is time-consuming and expensive. It is thus necessary to detect obsolete TODO comments and remove them automatically before they cause any unwanted side effects. 
In this work, we propose a novel model, named {\sc TDCleaner} (\textbf{\underline{T}}O\textbf{\underline{D}}O comment \textbf{\underline{Cleaner}}), to identify obsolete TODO comments in software projects. 
{\sc TDCleaner} can assist developers in just-in-time checking of TODO comments status and avoid leaving obsolete TODO comments.  
Our approach has two main stages: offline learning and online prediction. 
During offline learning, 
we first automatically establish $\langle code\_change, todo\_comment, commit\_msg \rangle$  training samples and leverage three neural encoders to capture the semantic features of  TODO comment, code change and commit message respectively. 
{\sc TDCleaner} then automatically learns the correlations and interactions between different encoders to estimate the final status of the TODO comment.  
For online prediction, we check a TODO comment's status by leveraging the offline trained model to judge the TODO comment's likelihood of being obsolete.
We built our dataset by collecting TODO comments from the top-10,000 Python and Java Github repositories and evaluated {\sc TDCleaner} on them.
Extensive experimental results show the promising performance of our model over a set of benchmarks.
We also performed an in-the-wild evaluation with real-world software projects,
we reported 18 obsolete TODO comments identified by {\sc TDCleaner} to Github developers and 9 of them have already been confirmed and removed by the developers,  demonstrating the practical usage of our approach.

\end{abstract}




\author{Zhipeng Gao}
\affiliation{%
\institution{Monash University}
\country{Australia}
}
\email{zhipeng.gao@monash.edu}

\author{Xin Xia}
\authornote{This is the corresponding author}
\affiliation{%
\institution{Monash University}
\country{Australia}
}  
\email{xin.xia@acm.org}

\author{David Lo}
\affiliation{%
\institution{Singapore Management University}
\country{Singapore}
}
\email{davidlo@smu.edu.sg}

\author{John Grundy}
\affiliation{%
\institution{Monash University}
\country{Australia}
}
\email{john.grundy@monash.edu}

\author{Thomas Zimmermann}
\affiliation{
\institution{Microsoft Research}
\country{United States}
}
\email{tzimmer@microsoft.com}



\maketitle

\section{INTRODUCTION}
\label{sec:intro}
\begin{figure}
\vspace{0.0cm}
\centerline{\includegraphics[width=0.50\textwidth]{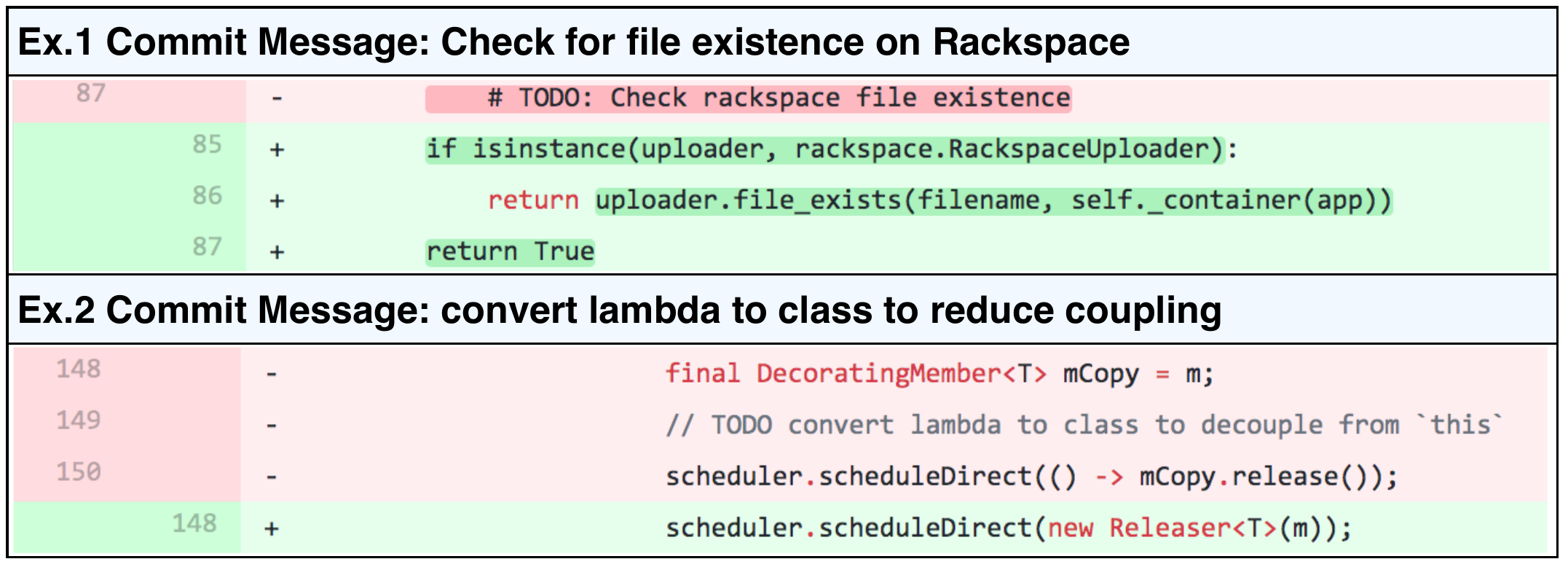}}
\caption{Example of code change and TODO comment}
\label{fig:intro_example}
\end{figure}

TODO comments in source code are extensively used by developers to denote their pending tasks. 
After completing (some of) the documented pending tasks, developers should update or remove their associated TODO comment(s). 
Such an example is shown in Ex.1 of Fig~\ref{fig:intro_example}. Here a developer added a TODO comment (Line 87, \emph{TODO: check rackspace file existence}) to notify themselves or others of the unfinished task. 
When the developer (or someone else in the team) updated the source code (highlighted in green colour) to perform this task, the accompanying TODO comment was also deleted (highlighted in red colour).
However, due to time constraints or carelessness, developers may have completed (or partially completed) the task specified by the TODO comment but forget to remove it (or update it)~\cite{tan2007icomment, wen2019large}.
This results in TODO comments becoming obsolete and more and more irrelevant and unreliable when the software changes and evolves. 


In this work, we define a TODO comment as an \emph{obsolete TODO comment} if its corresponding task is accomplished but the TODO comment itself is not removed. 
These obsolete TODO comments lay down outdated tasks that need not, or should not, be followed any longer.
A good TODO comment can help developers in understanding the designed tasks as well as the source code~\cite{de2005study, ying2005source}. 
On the contrary, an obsolete TODO comment can obscure the source code and affect code comprehension.  
Previous studies~\cite{tan2007icomment, ibrahim2012relationship, parnas2011precise} have shown that developers lose confidence in the reliability of the system when they encountered outdated comments.
Moreover, once obsolete TODO comments get separated from their code, they become incorrect and unreliable documentation of ever-decreasing accuracy, which can mislead developers and cause the introduction of bugs in the future~\cite{tan2007icomment, tan2011acomment}.
Therefore, the obsolete TODO comments increase the cost of software maintenance and development, which have negative impact to the quality and reliability of the system. 
It is thus highly desirable to have a tool that provides just-in-time automatic detection of obsolete TODO comments and removes them before they mislead developers and cause any damages.
However, making such a tool is difficult with respect to the following challenges:
\begin{itemize}
    \item {\em Capturing comment semantics} -- 
    Detecting obsolete TODO comments first requires understanding the semantics of the comments.
    Compared to source code, comments are written in natural language and have no mandatory format. 
    When source code changes, they can go through a series of software testing to ensure their correctness, however, there is no way to test comments to see if they are still valid or not. 
    Moreover, the source code and comments are of different types (together they form a heterogeneous data) 
    that cannot be easily matched to each other due to their lexical gaps. 
    Therefore, it is a non-trivial task to determine which TODO comments have been addressed and which ones have not. 
    
    \item {\em Capturing correlations} -- 
    It is very hard to determine if a TODO comment is resolved or not by just reading the source code.
    A more obvious and reliable way is to check the code change history with respect to the TODO comment, and determine if these code changes resolve the associated task, as shown in Ex.1 in Fig.~\ref{fig:intro_example}. 
    However, sometimes checking the code changes alone is not sufficient. 
    Ex.2 in Fig.~\ref{fig:intro_example} presents such a case. Even though the TODO comment and code change are presented, one can not easily claim this TODO comment is resolved due to his/her unfamiliarity with the code. The associated commit message can provide additional clues to fill this gap. 
    Therefore, to better identify TODO comments are up-to-date or obsolete, it is necessary to consider the correlations between the TODO comments, code changes and commit messages. 
\end{itemize}


In this work, 
to help developers better maintain the TODO comments in their software systems, we propose a novel neural network model, named {\sc \textbf{TDCleaner}} (\textbf{\underline{T}}O\textbf{\underline{D}}O comment \textbf{\underline{Cleaner}}), which can automatically detect the stale TODO comments in software repositories. 
{\sc TDCleaner} consists of two phases: offline learning and online prediction. 
During offline learning, we collect TODO comments from TOP-10,000 Python and Java Github repositories respectively.  
We automatically establish positive and negative  training samples in terms of whether a TODO comment is resolved or not. 
Our {\sc TDCleaner} can be trained as a binary classification model with these two kinds of training samples. 
To capture the semantics of the heterogeneous data, we employ three encoders, i.e., \textit{TODO Comment Encoder}, \textit{Code Change Encoder}, and \textit{Commit Message Encoder}, to embed TODO comments, code changes, and commit messages into contextualized vectors respectively. 
{\sc TDCleaner} then learns correlations and interactions between them by optimizing the final probability score. 
When it comes to online prediction, for a given TODO comment, we pair it with the associated code change and commit message, and fit them into the trained {\sc TDClearner} model to estimate their matching score. 

To verify the suitability of our proposed model, we conducted extensive experiments on Python and Java datasets. 
By comparing with several benchmarks, the superiority of our proposed {\sc TDCleaner} model is demonstrated.
In summary, this work makes the following main contributions:
\begin{enumerate}
    \item We propose a novel model, {\sc TDCleaner}, to automatically detect obsolete TODO comments by mining the histories of the software repositories. 
    {\sc TDCleaner} can help developers to increase the quality and reliability of software, and alleviate the error-prone code review process. 
    
    \item We build a large dataset for checking obsolete TODO comments from Github repositories, which contains more than 410K TODO comments for Python and more than 350K TODO comments for Java dataset. To the best of our knowledge, this is the first and by far the largest dataset for this task. 
    
    \item We extensively evaluate {\sc TDCleaner} using real-world popular open-source projects in Github. {\sc TDCleaner} is shown to outperform several baselines and reduce the developer's efforts in maintaining the TODO comments.  
    
    \item We have released our replication package~\cite{replicate}
    , including the dataset and the source code of {\sc TDCleaner}, to facilitate other researchers and practioners to repeat our work and verify their ideas. 
    
\end{enumerate}

The rest of the paper is organized as follows.
Section~\ref{sec:moti} presents the motivating examples and user scenarios of our study. 
Section~\ref{sec:approach} presents the details of our approach.
Section~\ref{sec:data} presents the data preparation for our approach. 
Section~\ref{sec:eval} presents the baseline methods , the evaluation metrics, and the evaluation results.
Section~\ref{sec:dis} presents the in-the-wild evaluation.
Section~\ref{sec:threats} presents the threats to validity. 
Section~\ref{sec:related} presents the related work.
Section~\ref{sec:con} concludes the paper with possible future work.

\section{MOTIVATION}
\label{sec:moti}

We show several motivating examples from popular Github repositories of the sorts of problems mentioned above. We then present user scenarios of employing our proposed approach to address these problems.

\subsection{Motivating Examples}

\begin{figure}
\centerline{\includegraphics[width=0.50\textwidth]{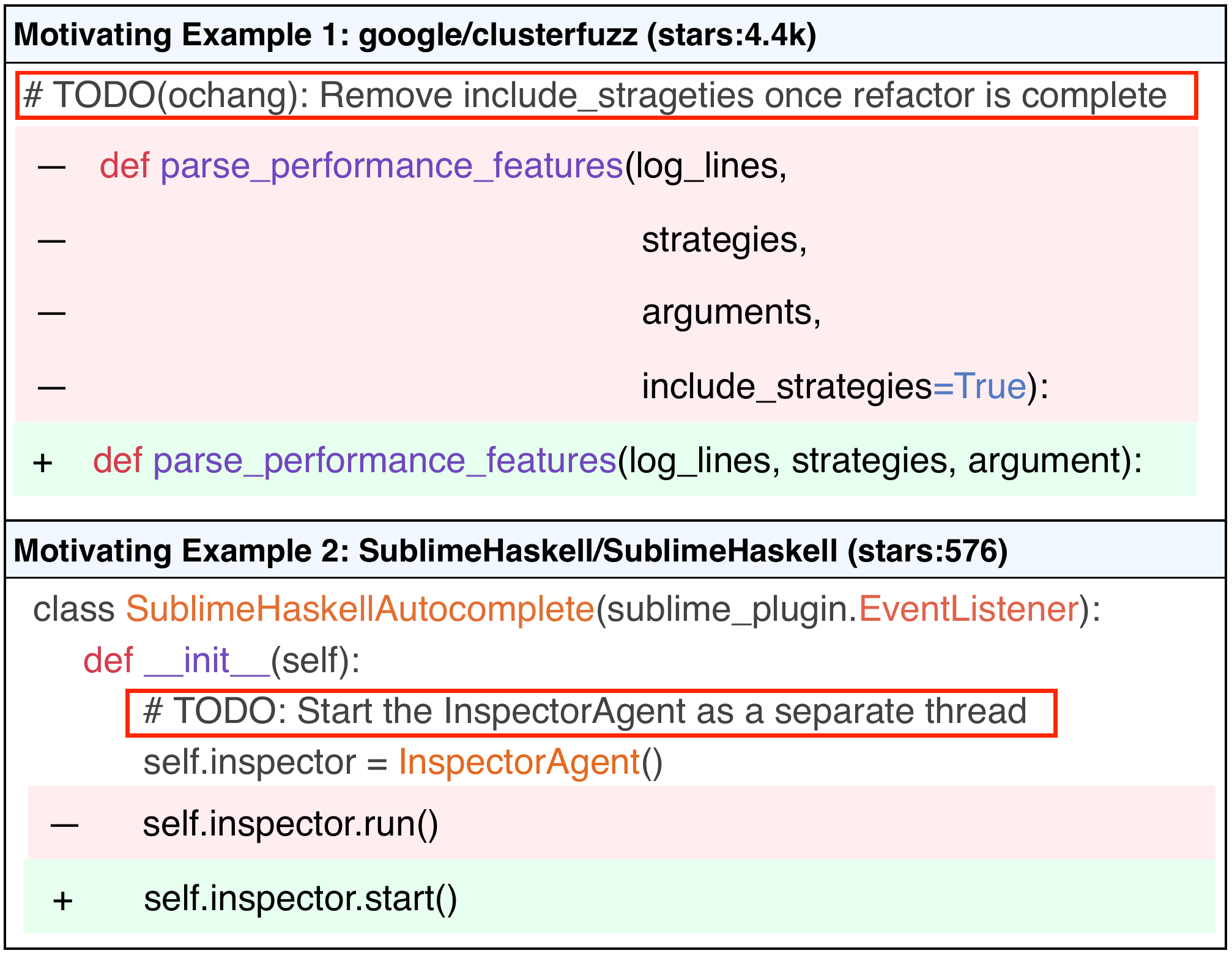}}
\caption{Motivating Examples}
\label{fig:moti_example}
\end{figure}

Developers change their source code but forget to remove or update associated TODO comments from time to time. 
Fig~\ref{fig:moti_example} shows two examples of stale TODO comments we found in real-world Github repositories. We can see that:
    
\textbf{Producing obsolete TODO comments is not only done by inexperienced developers:} Even developers within experienced teams (such as Ethereum and Google) may ignore or neglect such updates. 
For example, Motivating Example 1 presents such an obsolete TODO comment example from the \emph{clusterfuzz} project developed by the Google team. 
The TODO comment within the method \texttt{get\_client\_ip} states that ``\textit{remove include\_strategies once refactor is complete}''. 
However, when the task was resolved and \textit{include\_strategies} was removed from the method, the corresponding TODO comment was not updated.
Until the time we report this issue, this stale TODO comment has existed for over 1 year.
During this time, such a comment may hinder  program comprehension, cause miscommunication between developers, and confuse developers who perform the subsequent development. 
    
\textbf{A lot of software bugs are caused by the mismatch between code's implementation and developer's intention:}  
While a stale TODO comment itself might be harmless, it can mislead developers and cause the introduction of bugs in the future. 
For example, in our Motivating Example 2, the task TODO comment was fulfilled and should not be followed any longer. However, since the TODO comment stayed around, other developers can easily misunderstand the software component and violate the assumption and later lead to bugs. 

\subsection{User Scenarios}
We illustrate a usage scenario of {\sc TDCleaner} as follows:

\textbf{Without Our Tool:}
Consider a developer Bob. 
Daily, when Bob reads other people's code to perform the development, he encounters a few stale TODO comments. 
These out-of-date TODO comments can clutter the code and have a harmful effect for Bob to understand the current state of the code correctly.
Bob may lose confidence in the reliability of the system and even ignore the remainder of useful comments. 
Furthermore, since Bob has no idea that this TODO comment had already been resolved, Bob tries his best to perform the ``pending'' task with respect to this task comment by refactoring the code. 
However, due to his unfamiliarity with the code or the system, these updates are risky and very likely to cause the introduction of bugs in the future. 

\textbf{With Our Tool:}
Now consider Bob adopts our {\sc TDCleaner}. 
Before working on the software, Bob can use our tool to automatically check for the presence of stale TODO comments.
Moreover, {\sc TDCleaner} can identify the specific code change which resolved the TODO comment, thus helping to improve developer's confidence in verifying the obsolete TODO comments. 
With the help of our tool, Bob successfully removes the stale TODO comments in the current software repository efficiently and accurately, which can increase the reliability and maintenance of the system and decrease the likelihood of introducing bugs.

\section{OUR APPROACH}
\label{sec:approach}
\begin{figure*}
\centerline{\includegraphics[width=0.87\textwidth]{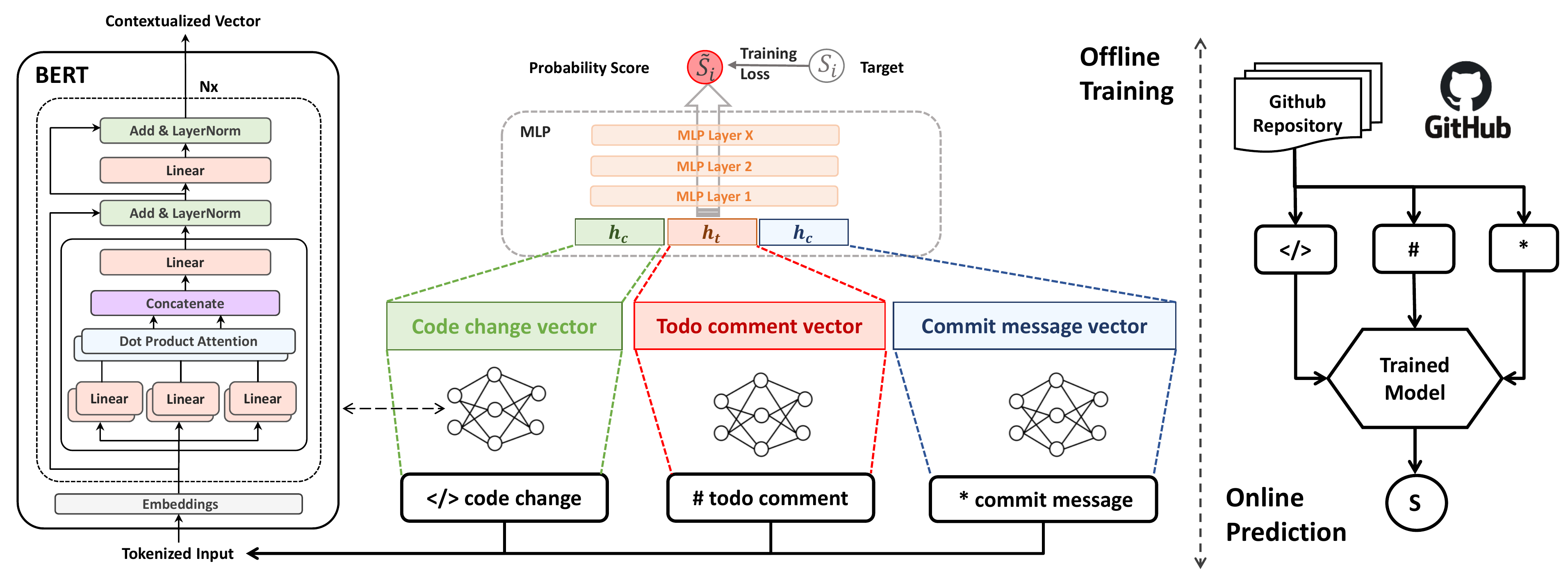}}
\caption{Overview of Our Approach}
\label{fig:workflow_approach}
\end{figure*}
 
We first define the task of identifying obsolete TODO comments for our study. 
We then present the details of our proposed model. 
The overall framework of our approach is illustrated in Fig.~\ref{fig:workflow_approach}. 

\subsection{Task Definition}
The motivation of our work is to automatically detect and remove obsolete TODO comments in software projects.
To do this, we need to detect if a TODO comment is resolved in a commit or not. We formulate this task as a binary classification learning problem as per below.
For a given commit, let $\mathbf{C}$ be the code changes extracted from the \texttt{diff} file, $\mathbf{M}$ be the corresponding commit message, 
and $\mathbf{T}$ be the TODO comment associated with the code changes (if a TODO comment appears within the default number of context lines of a code change, we claim this TODO comment is associated with the code change). 
Our target is to automatically determine the status $\mathbf{S}$ of the TODO comment, where $\mathbf{S}$ is positive when the TODO comment is resolved, and negative when the TODO comment is unresolved.  
In other words, our goal is to train a model $\theta$ using $\langle$$\mathbf{C}$, $\mathbf{T}$, $\mathbf{M}$$\rangle$ triples such that the probability $P_{\theta}(\mathbf{S}| \langle \mathbf{C},\mathbf{T},\mathbf{M} \rangle )$ is maximized over the given training dataset.
Mathematically, our task is defined as finding $\overline{y}$, such that:
\begin{equation}
    \overline{y} = argmax_{\mathbf{S}}P_{\theta}(\mathbf{S}| \langle \mathbf{C},\mathbf{T},\mathbf{M} \rangle)
\end{equation}
$P_{\theta}(\mathbf{S}| \langle \mathbf{C},\mathbf{T},\mathbf{M} \rangle )$ can be seen as the conditional likelihood of predicting the status $\mathbf{S}$ given the $\langle$$\mathbf{C}$, $\mathbf{T}$, $\mathbf{M}$$\rangle$ input triples.

\subsection{Approach Details}
\subsubsection{Encoders}
Identifying the resolved TODO comments from the unresolved ones is a non-trivial task. This is because the code changes and the TODO comments are of different types (source code vs. natural language) and cannot be easily mapped by simple matching of their lexical tokens.
To bridge this gap, {\sc TDCleaner} adopts three encoders, i.e., \textit{TODO Comment Encoder}, \textit{Code Change Encoder}, and \textit{Commit Message Encoder} to embed the code change, TODO comment and commit message into a vector representation respectively, so that semantically similar concepts across the three modalities can be correlated in the higher dimensional vector space. 
Through the embedding techniques, heterogeneous data can be easily linked through their vectors. 

Recently, neural networks have been widely used to capture the code semantic features by encoding code into vectors~\cite{gao2019smartembed, gao2020generating, gao2020checking, liu2020automating, hu2018deep, gu2018deep},
In this study, we employ BERT~\cite{devlin2018bert} as the encoder template for our task. 
This is because BERT has been proven to be effective for capturing semantics and context information of sentences in many other works~\cite{zhang2020sentiment, feng2020codebert, lewis2019bart, beltagy2019scibert, tabassum2020code}.
BERT consists of 12-layer transformers~\cite{vaswani2017attention}, each of the transformers being composed of a self-attention sub-layer with multiple attention heads. 
The input to each BERT embedding component is a sequence of tokens.
Given a sequence of tokens $\mathbf{x}=\{x_{1},...,x_{T}\}$ of length $T$ as input, BERT takes the tokens as input and calculate the contextualized representations $\mathbf{H}^{l}=\{h_{1}^{l},...,h_{T}^{l}\} \in \mathbb{R}^{T \times D}$ as output, where $l$ denotes the $l$-th transformer layer and $D$ denotes the dimension of the representation vector. 
We take the final hidden state of the first special token $\mathbf{h_{1}^{L}}$ as the embedding vector for the input sequence.
Our {\sc TDCleaner} consists of three encoders, i.e., \textit{Code Change Encoder}, \textit{TODO Comment Encoder} and \textit{Commit Message Encoder}.
These three encoders are nearly the same in structure and responsible for mapping three kinds of inputs, i.e., code changes, TODO comments and commit messages, into their corresponding embeddings.

\begin{itemize}
    \item \textit{Code Change Encoder:} The \textit{Code Change Encoder} embeds code changes into vectors. 
    Code changes contain multiple aspects of useful information such as code tokens, added lines and removed lines. 
    Consider a code change $\mathbf{C} = (C_1, C_2, ..., C_{N_{C}})$ comprising a number of $N_{C}$ tokens. 
    The \textit{Code Change Encoder} embeds it into a vector $\mathbf{h_{c}}$ using BERT. 
    
    \item \textit{TODO Comment Encoder:} 
    The \textit{TODO Comment Encoder} embeds the TODO comment into vectors. Consider a TODO comment $\mathbf{T} = (T_1, T_2, ..., T_{N_{T}})$ comprising a number of $N_{T}$ tokens. 
    After feeding $\mathbf{T}$ into the \textit{TODO Comment Encoder}, we can get the embedding vector $\mathbf{h_{t}}$ for the TODO comment.
    
    \item \textit{Commit Message Encoder:} 
    The \textit{Commit Message Encoder} embeds commit messages into vectors. Likewise, the commit message $\mathbf{M} = (M_1, M_2, ..., M_{N_{M}})$ is embeded into a contextualized vector $\mathbf{h_{m}}$. 
\end{itemize}

\subsubsection{Multi Layer Perceptron.}
So far, the code change, TODO comment and commit message are represented as independent contextual vectors. 
To capture the relationships between them, it is necessary to link and fuse their information. 
To do this, we add a Multi-Layer Perceptron (MLP)~\cite{he2017neural, gao2020technical} to address this need, which is shared with the three encoders.
We first concatenate three encoders' contextual vectors, i.e., $\mathbf{h_{c}}$, $\mathbf{h_{t}}$, and $\mathbf{h_{m}}$, to combine the semantic features.
To further capture the correlation and reference of the latent feature vectors, we next add a standard MLP on the concatenated vector. 
In this sense, we can endow the model with a large level of flexibility and non-linearity to learn the interactions between the three encoders. 
The MLP takes the the contextualized representations (i.e., $\mathbf{h_{c}}$, $\mathbf{h_{t}}$, and $\mathbf{h_{m}}$) as input and outputs the likelihood of the final status $\mathbf{S}=\{0, 1\}$.
More precisely, the MLP is defined as follows:
\begin{equation}
\label{eq:log_loss}
\begin{split}
&\mathbf{z}_{1} = \phi_{1} ( \mathbf{h}_{c}, \mathbf{h}_{t}, \mathbf{h}_{m} ) =  \left[ \begin{array}{l} 
\mathbf{h}_{c} \\ 
\mathbf{h}_{t} \\
\mathbf{h}_{m} \end{array}
\right] \\
&\mathbf{z}_{2} = \phi_{2} ( \mathbf{z}_{1} ) = \mathbf{a}_{2} ( \mathbf{W}_{2}^{T} \mathbf{z}_{1} + \mathbf{b}_{2} ) \\
&... \\
&\mathbf{z}_{L} = \phi_{L} ( \mathbf{z}_{L-1} ) = \mathbf{a}_{L} ( \mathbf{W}_{L}^{T} \mathbf{z}_{L-1} + \mathbf{b}_{L} ) \\
& P ( \mathbf{S} = j |\langle \mathbf{C}, \mathbf{T}, \mathbf{M} \rangle) = \sigma ( \mathbf{z}_{L} ) \\ 
\end{split}
\end{equation}

$\mathbf{W}_{x}$, $\mathbf{b}_{x}$, and $\mathbf{a}_{x}$ denote the weight matrix, bias vector, and activation function for the $x$-layer's perceptron respectively. $\sigma$ is the sigmoid function $\sigma(x) = 1/(1+e^{-x})$ which will output the final probability of status $\mathbf{S}$ between 0 and 1. 
For the probability score, we want this score to be high if the TODO comment is resolved and to be low if the TODO comment is unresolved. 

\section{DATASET PREPARATION}
\label{sec:data}
We present the details of our data collection and construction process.
We build our dataset by collecting data from the top 10,000 repositories (ordered by the number of stars) in Github for Python and Java repositories.
To the best of our knowledge, this is the first and by far the largest dataset for collecting TODO comments from Github repositories. 

\subsection{Data Collection}
\subsubsection{Identifying TODO Related Commits}
We first clone the top 10,000 repositories from Github.
The git repository stores software update history, each update comprises a commit message alongside a \texttt{diff} that represents the differences between the current and previous version of the affected files. 
For each cloned repository, we first checkout all the commits from the repository history. 
Following that, for each commit, if ``TODO'' appears within the \texttt{diff}, we consider this commit as a \emph{TODO related commit}. 
We have identified more than 410K TODO related commits from our Python repositories and more than 350K commits from our Java repositories. 

\subsubsection{Normalize Diff and Commit Message}
\begin{figure}
\vspace{0.0cm}
\centerline{\includegraphics[width=0.50\textwidth]{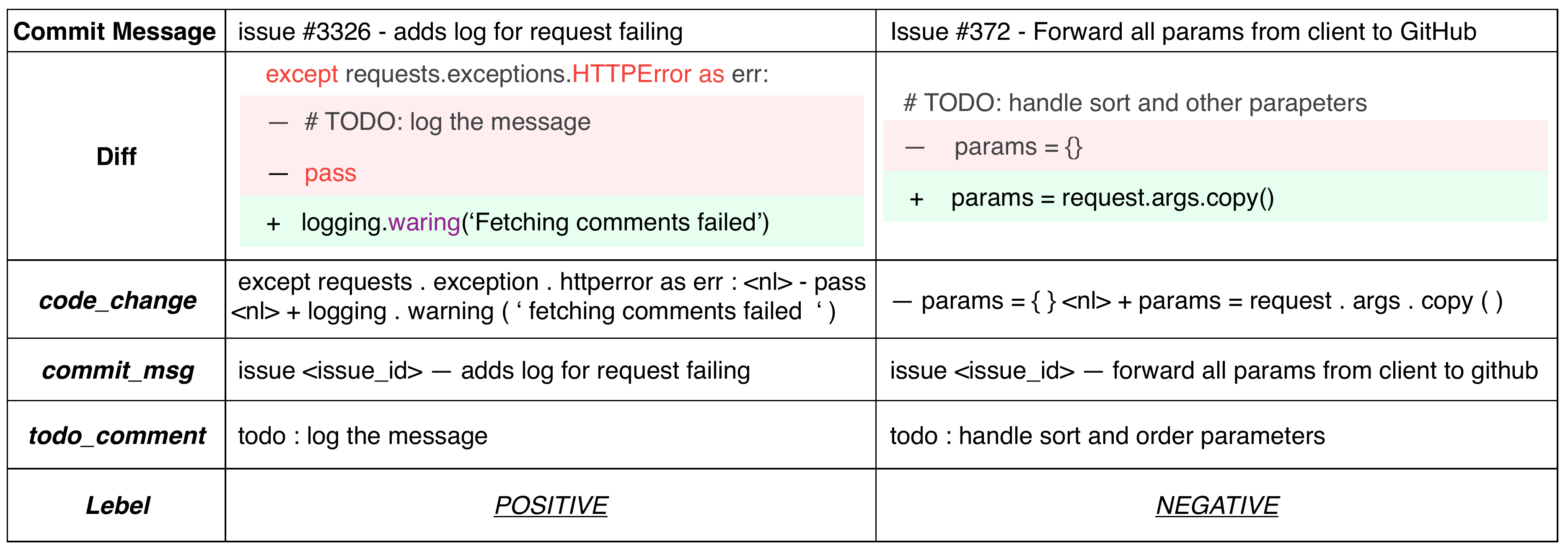}}
\vspace*{-0pt}
\caption{Data Preparation Examples}
\label{fig:data_pre_example}
\end{figure}

This step aims to normalize the commit sequence and remove some semantic-irrelevant information.
After identifying the TODO related commits, we extract the \texttt{diff} and the commit message from the commits and normalize them for later use. 
For the \texttt{diff}, we convert it into lowercase and delete the \texttt{diff} header by using regular expressions. 
Commit IDs within the \texttt{diff} are replaced with ``$\langle$commit\_id$\rangle$'' and commits with a \texttt{diff} larger than 1MB are removed. 
For the commit message, we first lowercase the commit message and retain the first sentences of the \emph{commit message} as the target, since the first sentences are usually the summaries of the entire commit message~\cite{gu2016deep, jiang2017automatically}.
Similar to \texttt{diff} normalization, Github issue IDs and commit IDs are replaced by ``$\langle$issue\_id$\rangle$'' and ``$\langle$commit\_id$\rangle$'' respectively to ensure semantic integrity.
The two examples in Fig.~\ref{fig:data_pre_example} demonstrates the process of normalization.

\subsubsection{Extract TODO Comments} 
This step is responsible for extracting TODO comments from the corresponding \texttt{diff}.  
A \texttt{diff} may contain multiple TODOs. We remove such instances because they are likely to be comment updates which might introduce noise for the later data construction process. 
Hereafter, each \texttt{diff} contains a single TODO comment and pairs with a commit message.  
We then use regular expressions to locate comments within the \texttt{diff} -- if ``TODO'' appears in the comment, then this comment is identified as a TODO comment. 
The TODO comment is extracted out of the \texttt{diff} as \textit{todo\_comment}, and the rest of the \texttt{diff} stay the same, referred to as \emph{code\_change} in this paper. 
It is worth mentioning that the difference label, i.e., ``+'' and ``-'', are deleted from the TODO comment, otherwise our model can learn directly from the difference labels instead of learning the semantics of the TODO comments. 
Until now, we are able to establish $\langle code\_change, todo\_comment, commit\_msg \rangle$ triples. 

\subsection{Data Construction}
\subsubsection{Data Labelling.}
\label{sec:data_label}
In this work, we consider a \texttt{diff} consists of three parts: \textit{LinesAdded}, \textit{LinesRemoved}, and \textit{LinesEqual}. 
\textit{LinesAdded} are the lines that were added to the current version when compared to its previous version. 
Similarly, \textit{LinesRemoved} are the lines that were removed from the current version when compared to its previous version.  \textit{LinesEqual} were the lines that remained unchanged between two versions. 
We automatically label each $\langle code\_change, \\ todo\_comment, commit\_msg \rangle$ triple as follows: 
(i) if \emph{todo\_comment} is within the scope of the \textit{LinesRemoved}, which means the TODO task is already performed by the developer, we label such triple instances as \textbf{Positive Samples}. 
(ii) if \emph{todo\_comment} is within the scope of the \textit{LinesEqual}, which means the \emph{code\_change} is not responsible for resolving the TODO task, we label such triple instances as \textbf{Negative Samples}. 
(iii) if \emph{todo\_comment} is within the scope of the \textit{LinesAdded}, we ignore such triple instances. 
This is reasonable since the \emph{todo\_comment} within the scope of \textit{LinesAdded} corresponds to the first time TODO comments are added.
However, our research focuses on detecting obsolete TODO comments from the existing ones. This requires that the TODO comments have been already added to the source code. 
Fig.~\ref{fig:data_pre_example} demonstrates the data labelling results for a positive sample and a negative sample respectively. 
We found more than 38K triple instances that are considered as positive samples and 35K triple instances that are considered as negative samples for Python, and 33K positive samples and 32K negative samples for Java. 

\subsubsection{Manual Checking}
We automatically built our positive and negative training samples via heuristic rules. 
We can not ensure that there are no outlier cases during the label establishment process. 
Therefore, we performed a manual checking step to examine the label of the training sample is correct or not. 
We randomly sampled 200 samples (including 100 positive samples and 100 negative samples) from our dataset. 
Then, the first author manually examined each sample and classified the \emph{todo\_comment} is resolved or not based on checking the associated \emph{code\_change} and \emph{commit\_msg}. 
Finally, 93 of the positive training samples are marked as resolved and all of the negative training samples are marked as unresolved. 
Thus, we are confident in the labels of our provided dataset.

\subsubsection{Data Splitting}
We split the constructed data samples into three chunks: 80 percent of the triple samples are used for training, 10 percent are used for validation and the rest are held-out for testing. 
The training set is used to adjust the parameters, while the validation set is used to minimize overfitting,
and the testing set is used only for testing the final solution to confirm the actual predictive power of our model with optimal parameters.
The number of training, validation, test sets of Python and Java dataset are displayed in Table~\ref{tab:data_sta}.

\begin{table} 
\caption{Data Statistics}
\begin{center}
\begin{tabular}{||c|l|c||}
    \hline  
    \multirow{5}{*}{Python}   & \# TODO Commits  & 416,666  \\ \cline{2-3}
                              & \# Positive samples  & 38,175 \\ \cline{2-3}
                              & \# Negative samples  & 35,995 \\ \cline{2-3}
                              & \# Train Set & 59,336 \\ \cline{2-3}
                              & \# Val\&Test Set & 7,417 \\ \cline{2-3}
    \hline
    \multirow{5}{*}{Java}   & \# TODO Commits  & 351,266  \\ \cline{2-3}
                          & \# Positive samples  & 33,797 \\ \cline{2-3}
                          & \# Negative samples  & 32,490 \\ \cline{2-3}
                          & \# Train Set & 53,029 \\ \cline{2-3}
                          & \# Val\&Test Set & 6,629 \\ \cline{2-3}
    \hline  
\end{tabular}
\label{tab:data_sta}
\end{center}
\end{table}

\section{Empirical Evaluation}
\label{sec:eval}
We first present the baselines, the evaluation metrics and our experiment settings. 
We then describe the results of our quantitative evaluation and manual analysis.

\subsection{Baselines} 
To demonstrate the effectiveness of our proposed model, {\sc TDCleaner}, we compared it with the following chosen baselines:
\begin{itemize}
    
    \item \textbf{TCO:}  \textbf{T}ODO-\textbf{C}odeChange-\textbf{O}verlap is a reasonable heuristic baseline to identify if a TODO task comment is resolved or not. TCO looks at the overlapping words in the \emph{todo\_comment} and \emph{code\_change}. 
    For example, as shown in the positive example in ~\ref{fig:data_pre_example}, the \textit{logging} function was added to the source code to resolve the TODO comment ``\textit{TODO: log the message}''. 
    Therefore, we created the \textbf{TCO} baseline as follows: if there is a common word between the \emph{todo\_comment} and the \emph{code\_change}, then we declare this TODO task comment to be resolved. 
    
    \item \textbf{TMO:}  \textbf{T}ODO-Commit\textbf{M}sg-\textbf{O}verlap baseline is based on the overlapping words of the \emph{todo\_comment} and the \emph{commit\_msg}. 
    In general, commit messages are meant to explain the purpose of the source code changes. We thus make the \textbf{TMO} baseline as follows: we compare the words in the \emph{todo\_comment} and the \emph{commit\_msg}, if there is a match we claim this task TODO comment to be resolved. 
    
    \item \textbf{TCMO:} We combined the status predicted by \textbf{TCO} and \textbf{TMO} to make the \textbf{TCMO} baseline.  
    For a testing sample, we declare this task TODO comment as resolved if either \textbf{TCO} or \textbf{TMO} method predicts so. 
    
    \item \textbf{IRSC:} Sridhara~\cite{sridhara2016automatically} proposed a method, \textbf{IRSC} (\textbf{I}nformation \textbf{R}etrieval Based \textbf{S}tatus \textbf{C}hecker), that performs well in identifying the status of the task TODO comments. 
    Different from our task, their method requires the signature and body of a method as input, then automatically checks if the associated task comment is up date with the code or obsolete.
    \textbf{IRSC} uses a cosine similarity between the task comment and the \textit{LinesAdded} by incorporating TF-IDF weightings. 
\end{itemize}

\subsection{Evaluation Metrics}
We define four statistics with respect to our task: 
(i) True Positive (TP) represents the number of resolved TODO comments that are classified as resolved. 
(ii) True Negative (TN) represents the number of unresolved TODO comments that are classified as unresolved. 
(iii) False Positive (FP) represents the number of unresolved TODO comments that are classified as resolved. 
(iv) False Negative (FN) represents the number of resolved TODO comments that are classified as unsolved. 
Based on the aforementioned four statistics, we adopted the widely-accepted metrics, i.e., Accuracy, Precision, Recall, and F1-score to evaluate the performance of {\sc TDCleaner} and baselines. 

Our evaluation metrics are defined as follows: 
\begin{itemize}
    
    \item \textbf{Accuracy:} Accuracy represents the proportion of correct predictions (both true positives and true negatives) among the total number of cases examined. 
    The Accuracy metric is defined as follows: 
    
    \begin{equation}
    Accuracy = \frac{TP + TN}{TP + TN + FP + FN}
    \end{equation}

    \item \textbf{Precision:} Precision represents the proportion of TODO task comments that are correctly classified resolved among all the resolved comments. 
    The Precision metric is defined as follows: 
    
    \begin{equation}
    Precision = \frac{TP}{TP + FP}
    \end{equation}
    
    \item \textbf{Recall:} Recall represents the proportion of all resolved task comments that are correctly classified as resolved. The Recall metric is defined as follows: 
    
    \begin{equation}
    Recall = \frac{TP}{TP + FN}
    \end{equation}
    
    \item \textbf{F1-score:} F1-score is the harmonic mean of precision and recall, which  combines both of the two metrics above. 
    It evaluates if an increase in precision (or recall) outweighs a reduction in recall (or precision), respectively. 
    The F1-score metric is defined as follows: 
    
    \begin{equation}
    F1 = \frac{2 \times Precision \times Recall} {Precision + Recall}
    \end{equation}
\end{itemize}

The higher an evaluation metric, the better a method performs. Note that there is a trade off between Precision and Recall. F1-score provides a balanced view of precision and recall.

\subsection{Experimental Settings}
We implemented {\sc TDCleaner} in Python using the Pytorch framework. 
We used the pre-trained BERT model~\cite{devlin2018bert} as the encoder for embedding training samples, which provides a powerful context-dependent sentence representation. 
BERT can be easily extended to a joint classification model.
In our model, \textit{Code Change Encoder}, \textit{TODO Comment Encoder} and the \textit{Commit Message Encoder} are jointly trained to minimize the cross entropy. 
After the encoding process, \textit{code\_change}, \textit{todo\_comment} and \textit{commit\_msg} will be mapped to a 768 dimensional vector respectively. 
During the training phase, we optimized the parameters of our model (including the BERT parameters and MLP parameters) using Adam~\cite{kingma2014adam} with a batch size of 32.
We use the ReLu as the activation function and employ three hidden layers for MLP. 
A dropout~\cite{srivastava2014dropout} of 0.2 is used for dense layers before computing the final probability. 
The model is validated every 1,000 batches on the validation set with a batch size of 32. 
We set the learning rate of Adam to 0.001 and clip the gradients norm by 2. 
The model with the best performance on the validation set was used for our evaluations. 

\subsection{Quantitative Analysis}
\label{subsec:quantitative_eval}
\subsubsection{RQ1: The Effectiveness Evaluation}
To evaluate the effectiveness of our proposed model, i.e., {\sc TDCleaner}, we evaluate it and the baseline methods on our testing set in terms of Accuracy, Precision, Recall and F1-score.
The evaluation results for Python and Java dataset are shown in Table~\ref{tab:effective_eval_python} and Table~\ref{tab:effective_eval_java} respectively. 
From the table, we can observe the following points:
\begin{itemize}
    \item In general, the methods based on bag-of-words matching, i.e., \textbf{TCO} and \textbf{TMO}, achieve the worst performance. 
    \textbf{TCO} and \textbf{TMO} methods identify the resolved TODO comments based on whether common words can be found in the input.
    As a result, they are unable to consider the context of the \textit{todo\_comment}, \textit{code\_change}, and \textit{commit\_msg} and the relationship between them, which is reflecting that simply checking the overlap words is not enough for our task. 
    It is notable that the \textbf{TMO} method gets relatively high Recall, i.e., 75.5\% for Python and 73.3\% for Java. 
    This is reasonable because \textit{commit\_msg} often describes the purpose behind commits, if \textit{commit\_msg} matches the \textit{todo\_comment}, it is very likely that the developer has completed this task comment and described this update in the \textit{commit\_msg}, an example is shown in the positive example of Fig.~\ref{fig:data_pre_example}.
    However, not all commit messages are meaningful and related to the task comment. This also explains its surprisingly low score on Precision and F1 score.
    
    \item \textbf{TCMO} performs better  than  \textbf{TCO} and \textbf{TMO} respectively. 
    For example, it improves over \textbf{TCO} on F1-score by 12.8\% on Python dataset and Java dataset, and it improves over \textbf{TMO} on F1-score by 62.1\% on Python dataset and 66.8\% on Java dataset. 
    Instead of solely based on \textit{code\_change} or \textit{commit\_msg}, 
    \textbf{TCMO} combines the advantage of \textbf{TCO} and \textbf{TMO} by incorporating useful information in both \textit{code\_change} and \textit{commit\_msg}. 
    This results in its superior performance to the other two approaches. It also signals that \textit{code\_change} and \textit{commit\_msg} convey much valuable information for our task of identifying the obsolete TODO comment.
    
    \item \textbf{IRSC} has its advantage as compared to the words overlapping based methods, i.e., \textbf{TCO}, \textbf{TMO} and \textbf{TCMO}. 
    Rather than checking if the words are overlapping, \textbf{IRSC} employed the TF-IDF metric, which can extract descriptive terms and identify the up-to-date status of the TODO comment by computing their similarities. 
    However, TF-IDF is still based on bag-of-words model, therefore \textbf{IRSC} can only capture the lexical level features, but unable to capture the semantic features and co-occurrences in separate input sequences.
    That is why its performance is comparatively suboptimal.
     
    \item \textbf{Our new model, i.e., {\sc TDCleaner}, outperforms all the baseline methods by a large margin} in terms of all evaluation metrics. 
    We attribute this to the following reasons: 
    First, it uses $\langle code\_change, todo\_comment, commit\_msg \rangle$ triple as input which considers the useful information across different resources. 
    Second, compared with bag-of-words models, the advantage of our proposed model is also clear.
    Our model employs  \textbf{BERT} as encoders to embed \textit{code\_change}, \textit{todo\_comment}, \textit{commit\_msg} into high dimensional vectors. These vector representations learn the semantic and structural features from the input and assists {\sc TDCleaner} to separate resolved task comments from the unresolved ones. 
    
    \item By comparing the evaluation results of the different datasets, i.e., Python and Java, we can see that {\sc TDCleaner} is stably and substantially better than the other baselines.
    This suggests that our approach behaves consistently across different programming languages. 
    This supports the likely generalization and robustness of our approach, which also justifies that our approach is language-independent and we believe it can be easily adapted to other programming languages. 
\end{itemize}

\begin{table}
\caption{Effectiveness Evaluation (Python)}
\label{tab:effective_eval_python}
\begin{center}
\begin{tabular}{|c|c|c|c|c|}
    \hline
    {\bf Measure} & {\bf Accuracy} & {\bf Precision} & {\bf Recall} & {\bf F1} \\
    \hline\hline
    \textbf{TCO}  & $55.3\%$ & $45.3\%$ & $58.1\%$ & $50.9\%$ \\
    \hline
    \textbf{TMO}  & $56.9\%$ & $23.1\%$ & $75.5\%$ & $35.4\%$ \\
    \hline
    \textbf{TCMO}  & $58.4\%$ & $54.8\%$ & $60.2\%$ & $57.4\%$ \\
    \hline
    \textbf{IRSC}  & $60.4\%$ & $61.0\%$ & $62.1\%$ & $61.6\%$ \\
    \hline
    {\sc \textbf{TDCleaner}}  & $\mathbf{84.7\%}$ & $\textbf{82.6\%}$ & $\mathbf{86.8\%}$ & $\mathbf{84.7\%}$ \\
    \hline
\end{tabular}
\end{center}
\end{table}

\begin{table}
\caption{Effectiveness Evaluation (Java)}
\label{tab:effective_eval_java}
\begin{center}
\begin{tabular}{|c|c|c|c|c|}
    \hline
    {\bf Measure} & {\bf Accuracy} & {\bf Precision} & {\bf Recall} & {\bf F1} \\
    \hline\hline
    \textbf{TCO}  & $56.5\%$ & $47.5\%$ & $58.7\%$ & $52.5\%$ \\
    \hline
    \textbf{TMO}  & $56.9\%$ & $23.4\%$ & $73.3\%$ & $35.5\%$ \\
    \hline
    \textbf{TCMO}  & $59.7\%$ & $57.6\%$ & $60.8\%$ & $59.2\%$ \\
    \hline
    \textbf{IRSC}  & $60.1\%$ & $59.0\%$ & $70.1\%$ & $64.0\%$ \\
    \hline
    {\sc \textbf{TDCleaner}}  & $\mathbf{85.0\%}$ & $\textbf{86.2\%}$ & $\mathbf{84.4\%}$ & $\mathbf{85.3\%}$ \\
    \hline
\end{tabular}
\end{center}
\end{table}

\vspace{5pt}
\noindent
\framebox{\parbox{\dimexpr\linewidth-2\fboxsep-2\fboxrule}{
\textbf{Answer to RQ-1: 
How effective is our approach for identifying the obsolete TODO comments? -- 
we conclude that our approach is highly effective for identifying the resolved TODO comments from the unresolved ones.}}} 
\vspace{5pt}

\subsubsection{RQ2: Component-Wise Evaluation} 
The key to our obsolete TODO detection task is to effectively capture the relationship and references between code changes and TODO comments. 
To do so, we adopt three encoders, i.e., \textit{Code Change Encoder}, \textit{TODO Comment Encoder}, and  \textit{Commit Mesage Encoder} to better represent and link information between \textit{code\_change}, \textit{todo\_comment} and \textit{commit\_msg}.  
To verify the effectiveness of these three encoders, we conduct a component-wise evaluation to evaluate their individual performance as well as their contributions one by one. 
We compare {\sc TDCleaner} with three of its incomplete versions: 
\begin{enumerate}

\item \vspace{0.1cm}\noindent {\bf TD\_CC\_Encoder:} it keeps the \textit{TODO Comment Encoder} and the \textit{Code Change Encoder}. It does not consider \textit{Commit Message Encoder} for this model. 
It is then trained as a binary classification model by using $\langle code\_change, todo\_comment \rangle$ pairs as the input. 

\item \vspace{0.1cm}\noindent {\bf TD\_MSG\_Encoder:} it keeps the \textit{TODO Comment Encoder} and the \textit{Commit Message Encoder}. It does not consider \textit{Code Change Encoder} for this model. 
{\bf TD\_MSG\_Encoder} can be trained with $\langle  \textit{todo\_comment}, \textit{commit\_msg} \rangle$ pairs same as above. 

\item \vspace{0.1cm}\noindent {\bf CC\_MSG\_Encoder:} it keeps the \textit{Code Change Encoder} and \textit{Commit Message Encoder}. It ignores the \textit{TODO Comment Encoder} for this model. 
Likewise, {\bf CC\_MSG\_Encoder} is trained with \textit{code\_change} and \textit{commit\_message} as input the same with above.

\item \vspace{0.1cm}\noindent {\bf {\sc TDCleaner}:} our model which contains all three Encoders. 
\end{enumerate}

The evaluation results are shown in Table~\ref{tab:component_eval_python} and Table~\ref{tab:component_eval_java} for Python and Java respectively. It can be seen that: 

\begin{itemize}
    \item \textbf{No matter which component we removed, it hurts the overall performance of our model. }
    This verifies our assumption that all the three encoders embed useful information from their input respectively. 
    
    \item The performance of \textbf{\textbf{TD\_CC\_Encoder} is better than the other two variants}. In other words, keeping the \textit{TODO Comment Encoder} and the \textit{Code Change Encoder} achieve a minimal performance drop. This justifies the importance and necessity of the above two encoders. 
    
    \item 
    \textbf{\textbf{TD\_MSG\_Encoder} achieves the worst performance.}
    It is clear there is a significant drop overall in every evaluation measure after removing the \textit{Code Change Encoder}. 
    This signals that the \textit{Code Change Encoder} is the most important of all the three encoders and has major contributions to the overall performance. 
    
    \item 
    \textbf{Even though \textbf{TD\_CC\_Encoder} don't get top results as {\sc TDCleaner}, it still achieves considerable performance} (e.g., with F1 score close to 80\%), which further confirms the strength of our approach.
    Moreover, since \textbf{TD\_CC\_Encoder} does not rely on the commit messages, it can be used as a light variant of our approach, which can help developers to remove the obsolete TODO comment just-in-time when code change happens. 

\end{itemize}

\begin{table}
\caption{Component-Wise Evaluation (Python)}
\label{tab:component_eval_python}
\vspace*{-10pt}
\begin{center}
\begin{tabular}{|l|c|c|c|c|}
    \hline
    {\bf Measure} & {\bf Accuracy} & {\bf Precision} & {\bf Recall} & {\bf F1} \\
    \hline\hline
    \textbf{TD\_CC\_Encoder}  & $78.4\%$ & $78.5\%$ & $79.0\%$ & $78.8\%$ \\
    \hline
    \textbf{TD\_MSG\_Encoder}  & $61.9\%$ & $62.0\%$ & $63.0\%$ & $62.5\%$ \\
    \hline
    \textbf{CC\_MSG\_Encoder}  & $76.1\%$ & $67.7\%$ & $82.4\%$ & $74.3\%$ \\
    \hline
    {\sc \textbf{TDCleaner}}  & $\mathbf{84.7\%}$ & $\mathbf{82.6\%}$ & $\mathbf{86.8\%}$ & $\mathbf{84.7\%}$ \\
    \hline
\end{tabular}
\end{center}
\end{table}

\begin{table}
\caption{Component-Wise Evaluation (Java)}
\label{tab:component_eval_java}
\vspace*{-10pt}
\begin{center}
\begin{tabular}{|l|c|c|c|c|}
    \hline
    {\bf Measure} & {\bf Accuracy} & {\bf Precision} & {\bf Recall} & {\bf F1} \\
    \hline\hline
    \textbf{TD\_CC\_Encoder}  & $78.6\%$ & $82.6\%$ & $76.9\%$ & $79.7\%$ \\
    \hline
    \textbf{TD\_MSG\_Encoder}  & $60.9\%$ & $59.6\%$ & $61.8\%$ & $60.7\%$ \\
    \hline
    \textbf{CC\_MSG\_Encoder}  & $77.0\%$ & $81.3\%$ & $75.3\%$ & $78.2\%$ \\
    \hline
    {\sc \textbf{TDCleaner}}  & $\mathbf{85.0\%}$ & $\mathbf{86.2\%}$ & $\mathbf{84.4\%}$ & $\mathbf{85.3\%}$ \\
    \hline
\end{tabular}
\end{center}
\end{table}

\vspace{5pt}
\noindent
\framebox{\parbox{\dimexpr\linewidth-2\fboxsep-2\fboxrule}{
\textbf{Answer to RQ-2: 
How effective is our use of \textit{TODO Comment Encoder}, \textit{Code Change Encoder} and \textit{Commit Message Encoder} under automatic evaluation? -- 
we conclude that all the three encoders are effective and helpful to enhance the performance of our model.}}} 
\vspace{5pt}

\subsubsection{RQ3: Ablation Evaluation} 
Another advantage of {\sc TDCleaner} is using a pre-trained BERT model to learn semantic features from the input.
The BERT model provides a powerful context-dependent sentence representation and has achieved the state-of-the-art performance on various NLP tasks, i.e., question answering, language inference, etc.
Therefore, we conduct an ablation analysis to verify the effectiveness of using BERT as embedding techniques for our task. 
In this research question, we compare our proposed model with \textbf{Drop-BERT}. 
\textbf{Drop-BERT} drops BERT from our model and does not use the embeddings pre-trained by BERT. Instead we replaced them with a traditional Word2Vec embedding technique.  

The results of our comparisons are presented in Table~\ref{tab:ab_eval}. 
By comparing the results of \textbf{Drop-BERT} and \textbf{Ours}, we can measure the performance improvement achieved by employing the BERT embedding techniques. 
\textbf{It is clear that dropping the BERT component  hurts the performance of our model}. 
For example, regarding the F1 score, the improvements achieved by adding BERT embeddings range from 25.2\% to 29.1\%. 
These results indicate that the BERT embeddings are useful and effective for our task of identifying the obsolete TODO comments. 

\vspace{5pt}
\noindent
\framebox{\parbox{\dimexpr\linewidth-2\fboxsep-2\fboxrule}{
\textbf{Answer to RQ-3: 
How effective is our approach for using BERT embedding techniques? -- 
we conclude that BERT model significantly improves the overall performance of our model.}}} 
\vspace{3pt}

\begin{table}
\caption{Ablation Evaluation}
\label{tab:ab_eval}
\vspace*{-10pt}
\begin{center}
\begin{tabular}{|l||c|c||c|c|}
    \hline
    \multirow{2}{*}{Measure} & \multicolumn{2}{c||}{Python} & \multicolumn{2}{c|}{Java} \\\cline{2-5} 
    & {\bf Drop-BERT} & {\bf Ours} & {\bf Drop-BERT} & {\bf Ours} \\
    \hline\hline
    \textbf{Accuracy} & $64.6\%$ & $\mathbf{84.7\%}$ & $66.2\%$ & $\mathbf{85.0\%}$ \\
    \hline
    \textbf{Precision} & $65.2\%$ & $\mathbf{86.2\%}$ & $65.2\%$ & $\mathbf{86.2\%}$ \\
    \hline
    \textbf{Recall} & $65.9\%$ & $\mathbf{86.8\%}$ & $71.1\%$ & $\mathbf{84.4\%}$ \\
    \hline
    \textbf{F1} & $65.6\%$ & $\mathbf{84.7\%}$ & $68.1\%$ & $\mathbf{85.3\%}$ \\
    \hline
\end{tabular}
\end{center}
\vspace{-6pt}
\end{table}

\subsection{Manual Analysis}
\begin{figure}
\vspace{0.0cm}
\centerline{\includegraphics[width=0.49\textwidth]{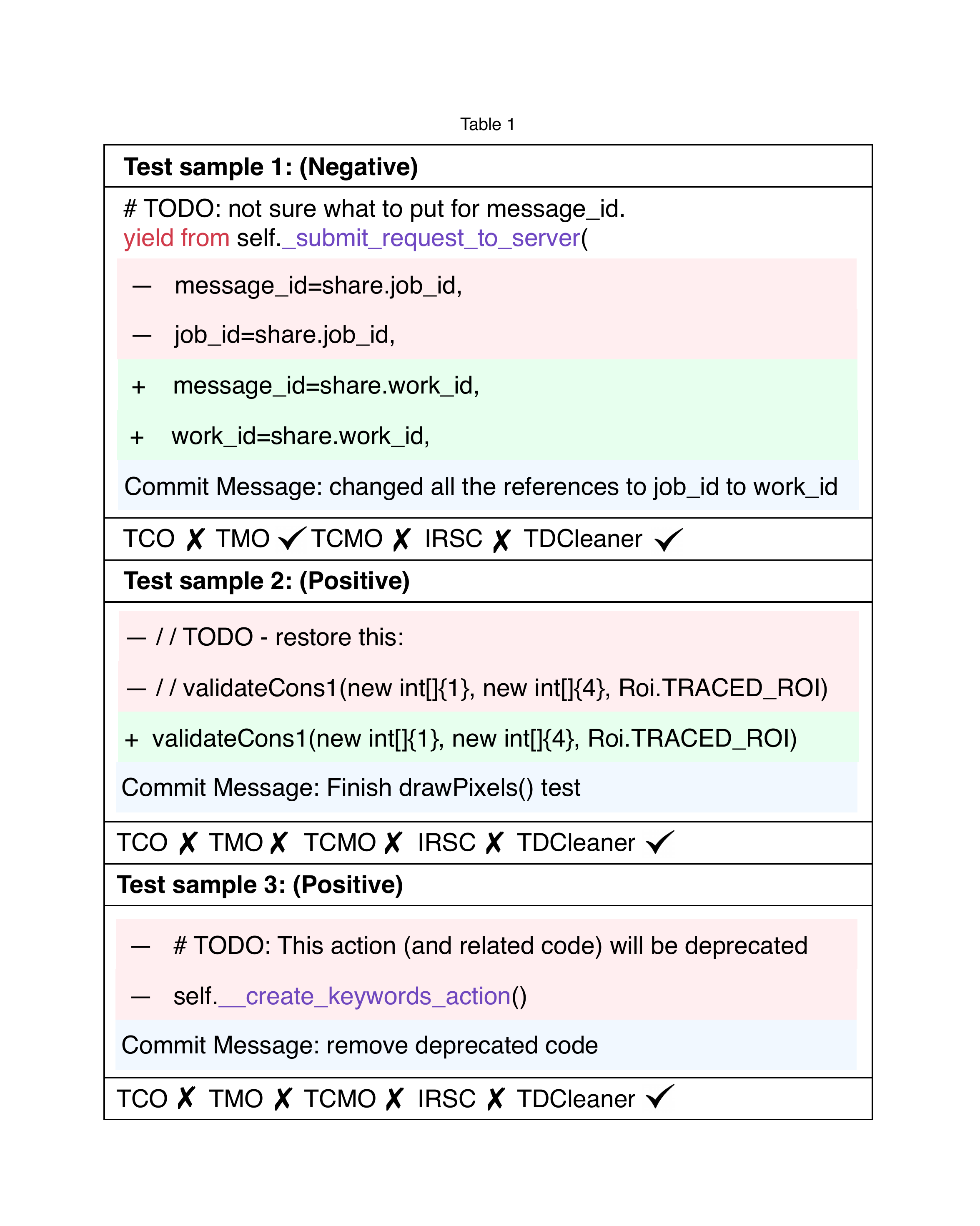}}
\caption{Manual Analysis Examples}
\label{fig:manual_example}
\end{figure}

To better understand the reasons why {\sc TDCleaner} outperforms other approaches, we manually inspected the test results. 
Based on our inspection, we summarize two major advantages of {\sc TDCleaner} as compared to other baseline approaches:

First, {\sc TDCleaner} automatically learns important semantic-level features from the vector representations, while the bag-of-words based models are limited to learn the lexical-level features. 
All of the baseline approaches, i.e., \textbf{TCO}, \textbf{TMO}, \textbf{TCMO} and \textbf{IRSC} rely on the lexical similarities between the input sequences, they are unable to consider semantic-level features. 
In contrast, our {\sc TDCleaner} leverages a probabilistic model to learn semantic features and capture common patterns from a very large scale of $\langle code\_change, todo\_comment, commit\_msg \rangle$ training samples. 
These patterns learned by {\sc TDCleaner} can cover more and diverse samples. 
For example, as shown in the test sample 2 of Fig.~\ref{fig:manual_example}, there is no similarity among words between the \textit{todo\_comment} and \textit{code\_change} or \textit{commit\_msg}, thus \textbf{TCO}, \textbf{TMO} and \textbf{TCMO} fail to output the correct prediction, while {\sc TDCleaner} successfully learns the semantics behind the TODO comment ``\textit{TODO - restore this}'', and therefore {\sc TDCleaner} detects this task is completed. 
Moreover, relying on words overlapping leads to unreliable and inaccurate results, as shown in test sample 1, the developer replaced \textit{job\_id} with \textit{work\_id}, however since the word ``\textit{message\_id}'' appeared in both \textit{todo\_comment} and \textit{code\_change}, \textbf{TCO}, \textbf{TCMO} and \textbf{IRSC} falsely identifies that this comment has been resolved. 

Second, {\sc TDCleaner} learns the correlations and references between the \textit{code\_change}, \textit{todo\_comment} and \textit{commit\_msg}. 
All the baselines do not take the co-evolution between code change and comment updates into consideration, and thus can not detect the interaction matching between them. 
Different from the baseline methods, {\sc TDCleaner} explicitly adopts three encoders and the MLP layer between them, which enables our model to effectively link the interaction relationships between \textit{todo\_comment} with \textit{code\_change} and \textit{commit\_msg}. 
Fig.~\ref{fig:manual_example} test sample 3 presents such an example,
all the baseline approach fail to handle such a case, however, our approach relates the task comment (``\textit{TODO: This action (and related code) will be deprecated}'') with the removed code lines and the commit message (``\textit{remove deprecated code}''), thus successfully identifies this TODO comment as resolved. 

We also inspected a number of samples where {\sc TDCleaner} failed to make correct predictions. 
These samples presented two common aspects that {\sc TDCleaner} is difficult to deal with. 
A common failed situation is that the code change or commit message do not provide sufficient information to judge the status of the TODO comment. 
For example, in a test sample, the developer removed the comment ``TODO: is the full url right?'' after adding a print statement ``print res.data''. 
However, since we lack contextual information of the data structure, {\sc TDCleaner} is not able to perform such a classification. 
Another situation is that the code changes are too complicated for {\sc TDCleaner} to learn. 
For example, to implement the TODO task, developers sometimes make many code changes across different sub modules. 
The complicated code changes can hinder our approach for capturing the code semantics.

\vspace{5pt}
\noindent
\framebox{\parbox{\dimexpr\linewidth-2\fboxsep-2\fboxrule}{
\textbf{Why Does Our Approach Succeed? -- 
we conclude that our approach automatically learns the correct semantics and correlations from the training samples.}}} 
\vspace{5pt} 


\section{In-the-wild evaluation}
\label{sec:dis}

Since the final goal of our approach is detecting and removing the obsolete TODO comments in software projects, we also perform an in-the-wild evaluation to evaluate the effectiveness of {\sc TDCleaner} for removing the stale TODO comments in real world software repositories in Github.

\begin{figure}
\vspace{0.0cm}
\centerline{\includegraphics[width=0.49\textwidth]{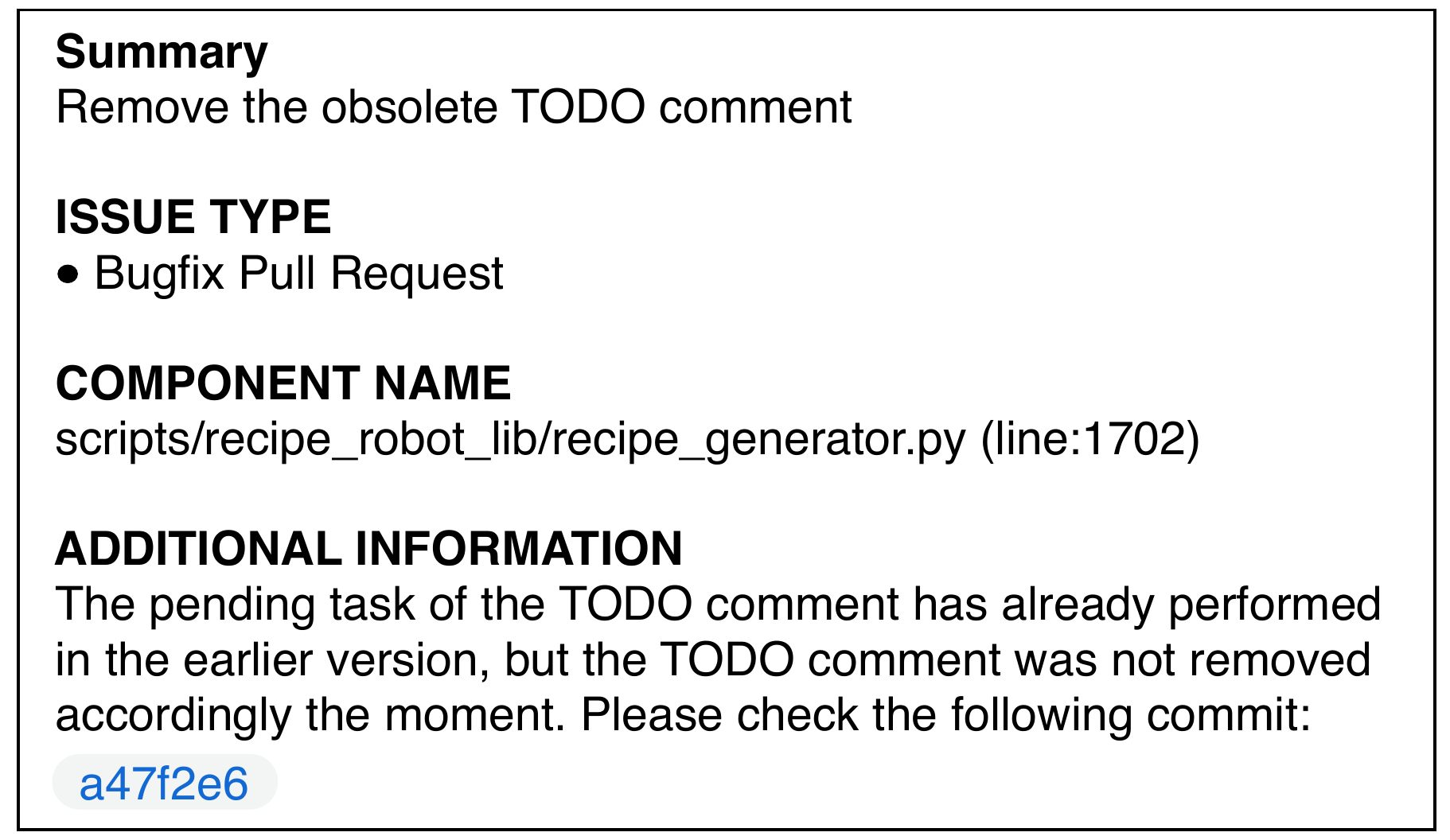}}
\caption{Pull Request Examples}
\label{fig:example}
\end{figure}

We randomly selected 100 Python repositories hosted on Github; for each Github repository, we first clone the repository from Github to local. 
Following that, we checkout all the commits of the repository to identify the TODO related commits. 
By going through the same steps of normalization, we extract the \textit{todo\_comment}, \textit{code\_change} and \textit{commit\_msg} from \texttt{diff} and commit messages.
We then build $\langle code\_change, todo\_comment, commit\_msg \rangle$ triple and feed the triple as input for {\sc TDCleaner}. 
For each constructed triple input,  
we apply {\sc TDCleaner} to check if the TODO comment has been resolved or not. 
If the TODO comment has been identified as resolved by our approach but not removed accordingly, we regard this TODO comment as obsolete. 

We ran {\sc TDCleaner} on these software projects and {\sc TDCleaner} reported that 23 TODO comments as being obsolete.
We manually checked all of these reported TODO comments, 5 of them are ``intermediate'' obsolete TODO comments, which means they
have been removed in subsequent versions. 
The reason for this phenomenon maybe that developers often clean up the obsolete TODO comments before specific software releases. 
Even though these TODO comments are finally removed during software development before being cleaned up by developers, they may mislead developers, waste their time to check the implementations, complicate the code review and cause the introduction of bugs.
It is necessary to remove the obsolete TODO comments just-in-time or avoid introducing them at all. 

In addition to the 5 ``intermediate'' obsolete TODO comments, there are still 18 ``potential'' obsolete TODO comments dangling in the current version of these projects. 
For each of these ``potential'' obsolete TODO comments, we would like to investigate if it is really an obsolete comment based on developer feedback.
To do this we submitted a pull-request to the corresponding Github repository by removing this ``potential'' obsolete TODO comment. 
To help developers verify the pull-request more easily and confidently, we also notified the developers for the commit for which our approach identifies this TODO comment has been resolved. 
An example of our submitted pull-request is shown in ~\ref{fig:practical_example}.
To avoid subjective bias, the developers don't know the pull-requests are automatically detected by our tool {\sc TDCleaner}. 
We submitted 18 pull-requests and 9 of them have already been confirmed and merged by the developers. 4 of them were closed by developers and the rest of them are still open. 
Fig.~\ref{fig:practical_example} demonstrates three examples we submitted to the developers.
From developers feedback to us we make the following observations:

\begin{figure}
\vspace{0.0cm}
\centerline{\includegraphics[width=0.49\textwidth]{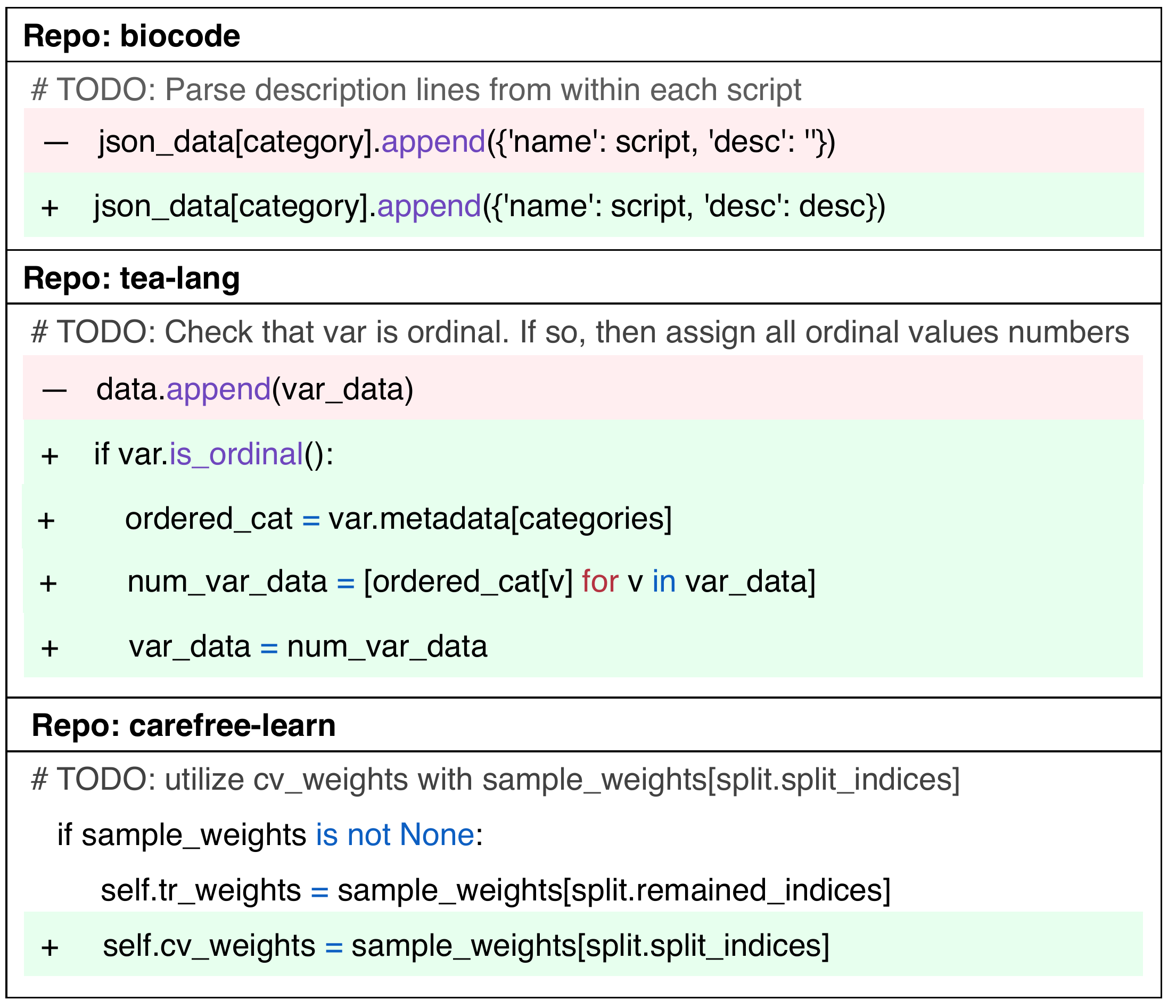}}
\caption{Practical Analysis Examples}
\label{fig:practical_example}
\end{figure}

\begin{itemize}
    \item For developers who have adopted our pull-request changes, they appreciated the contributions we made to their repositories.
    This signals that our proposed model, i.e., {\sc TDCleaner}, has positive effects to the robustness of the system and may decrease the cost of its development and maintenance.  
    
    \item Some developers expressed their curiosity about how we detected these obsolete TODO comments.
    For example, when we submitted pull-request to the first repository in Fig.~\ref{fig:practical_example}, a developer mentioned, \textit{``Indeed, thank you! I wonder how did you find this? Just browsing the code at leisure time? :)''}. 
    This is because this Github repository contains more than 4,000 commits, it is thus very hard, if not possible, to manually identify an obsolete TODO comment from the huge amount of commits accumulated in the repository.
    This also justifies the importance and necessity of developing a tool to automatically detect the obsolete TODO comments.  
    
    \item Instead of completely removing the obsolete TODO comments, some developers suggested another way of just removing the word ``TODO''. 
    The second example in Fig~\ref{fig:practical_example} presents such a situation. 
    When we submitted the pull-request to remove the obsolete TODO comment, the developer responded \textit{``How about just removing the word "TODO"? The rest of the comment is now good documentation.''}.
    This may be a reasonable approach because the TODO comments often describe the pending tasks that should be done, once the developer completed the task, the rest of the TODO could potentially serve as good documentations for the newly added code.  
    
    \item Not all the pull-requests we submitted are accepted by developers. The third case in Fig~\ref{fig:practical_example} shows such an example, the developer closed our pull-request and commented \textit{``Good catch! However although I've already recorded cv\_weights, I haven't actually utilized it yet. Which means, I haven't applied cv\_weights to get a weighted metrics on the validation set.''} 
    This indicates that sometimes the information is not enough for {\sc TDCleaner} to infer the correct prediction, for such cases, we still need developers to double check the results. 
    
\end{itemize}

\vspace{5pt}
\noindent
\framebox{\parbox{\dimexpr\linewidth-2\fboxsep-2\fboxrule}{
\textbf{
How effective is our {\sc TDCleaner} for removing obsolete TODO comments in real Github repositories? -- we conclude that our model is effective for detecting obsolete TODO comments in real word Github repos.}}} 
\vspace{5pt}

\section{Threats to Validity}
\label{sec:threats}


Several threats to validity are related to our research: In the data preparation process, we removed the diffs which contain multiple TODOs. 
When we evaluate our approach in practice, the evaluation data goes through the same data processing steps, which means the diffs with multiple TODOs are also removed before feeding them into our model. 
We have to admit our current approach lacks the ability of dealing with multiple TODOs when analyzing unseen data. 
We will try to address this shortcoming in our future work.  

Besides, our dataset is constructed from Python and Java projects in Github. 
This was because Python and Java are the most popular programming languages widely used by developers in GitHub. 
However, there are many other programming language projects in Github which were not considered in our study. 
Considering that our approach is language-independent, we argue that our approach can be easily adapted to other programming languages. 
We will try to extend our work to benefit more developers in the future. 

Regarding the model hyperparameters, there are two key hyperparameters for constructing our model, i.e., the embedding size of the encoders and the size of MLP hidden layers. 
Because we use the pre-trained BERT model as our encoders, the embedding size is fixed to 768 for BERT. 
Therefore, the only hyperparameter we can tune is the size of the hidden layers. 
Theoretically, we can fine-tune the size of hidden layers, however, due to the wide range of different size settings, the complexity of training is too expensive and time-consuming. We thus follow classic MLP layer settings~\cite{gao2020technical} in previous works and our approach has achieved promising results for the task.

Another threat to validity is that our work is focused on TODO comment in particular, which is a special case of Self-Admitted Technical Debt (SATD)~\cite{potdar2014exploratory}. 
Our preliminary study focuses on TODOs as opposed to other SATD, such as TODO, HACK, FIXME, etc. 
This is because a TODO often hints at the functionality that is not yet implemented, while HACK and FIXME might point to existing, but imperfect implementations. 
Therefore, we choose the obsolete TODO comments for our preliminary study, we plan to investigate the effectiveness of our approach to other types of STAD comments in the future.


\section{Related Work}
\label{sec:related}

\subsection{TODO Comments in SE Research}
TODO comments are widely used by developers to notify themselves or others of pending tasks. 
Prior works have investigated the role of TODO comments in software development and maintenance~\cite{storey2008todo, yan2018automating, potdar2014exploratory, huang2018identifying, sridhara2016automatically, nie2019framework, ren2019neural}. 

For example, Storey et al.~\cite{storey2008todo} performed an empirical study among developers to investigate how TODO comments are used and managed in software engineering.
They found that the use of task annotations varies from individuals to teams, and if incorrectly managed, they could negatively impact the maintenance of the system. 
Potdar and Shihab~\cite{potdar2014exploratory} proposed the self-admitted technical debt (SATD) concept (e.g., TODO, FIXME, HACK) for the first time, and found that 26.3-63.5\% of the SATD was removed after its introduction.
Huang et al.~\cite{huang2018identifying} used the text-mining based methods to predict whether a comment contains SATD or not. 
Rungroj et al.~\cite{maipradit2020wait} first introduced the concept of ``on-hold'' SATD and proposed a tool~\cite{maipradit2020automated} to identify and remove the ``on-hold'' SATD automatically.
Even if obsolete TODOs and ``on-hold'' SATD are similar, there are several differences between our work and theirs: (i) According to their definition in~\cite{maipradit2020wait}, the ``on-hold'' SATD contains a specific condition to be triggered, while the obsolete TODOs in our study do not have such constraint. 
(ii) their approach~\cite{maipradit2020automated} detects ``on-hold'' SATD based on the issue-referring comments, while our approach identifies general obsolete TODO comments based on commit histories. 

There is very limited work on detecting obsolete TODO comments.  
Sridhara~\cite{sridhara2016automatically} presented a technique for automatically identifying whether a TODO comment in a given method is up to date or obsolete. 
Different from their approach which requires the method's body and signature as input, our {\sc TDCleaner} takes the commits as input which is more general. 
Moreover, their approach is based on lexical level and heuristic rules defined by human experts, which can not capture the semantic features and/or adapt to different kinds of samples. 
We compared their approach, i.e., \textbf{IRSC} with {\sc TCleaner} in Section~\ref{subsec:quantitative_eval}, the evaluation results show that our approach outperforms theirs by a large margin. 

\subsection{Code-Comment Inconsistency Detection}
Inconsistent updates between code and comments are risky and should be carefully reviewed by practitioners.  
Previous studies~\cite{tan2007icomment, tan2007hotcomments, tan2011acomment, malik2008understanding, liu2018automatic, ratol2017detecting, panthaplackel2020learning, liu2020automating, zhong2013detecting, tan2012tcomment} have investigated the inconsistency between code and their documentations. 

Some works focused on the comments related to specific code properties or of specific types. 
For example, Tan et al.~\cite{tan2007icomment, tan2007hotcomments, tan2011acomment} proposed a series of approaches to detect code-comment inconsistency with respect to specific code properties, such as lock mechanisms, function calls and interrupts.
They use NLP techniques to extract the concept-related rules and use static program analysis to check source code against these rules. 
Some other works~\cite{malik2008understanding, liu2018automatic, ratol2017detecting} focused general comments and took code change into consideration. 
For example, Liu et al.~\cite{liu2018automatic} leveraged 64 manually-crafted features and machine learning techniques to check the code comment inconsistency. 
Most recently, Liu et al.~\cite{liu2020automating} employed a seq2seq model to automatically update bad comments in software projects. 
The techniques mentioned above check the inconsistency between code and comments. This is different from our work which is designed to detect obsolete TODO comments in software repositories. 


\section{Conclusion and Future work}
\label{sec:con}
This research aims to automatically detect and remove obsolete TODO comments from software repositories. 
To address this task, 
we first collect obsolete TODO comments from the top-10,000 Github repositories, 
To the best of our knowledge, this is the first and by far the largest dataset for detecting obsolete TODO comments. 
We propose an approach named {\sc TDCleaner} (\textbf{\underline{T}}O\textbf{\underline{D}}O comment \textbf{\underline{Cleaner}}), which leverage a neural network model to learn the semantic features and correlations between code changes, TODO comments and commit messages. 
Extensive experiments on the real-world Github repositories have demonstrated its effectiveness and promising performance.
In the future, we plan to investigate the effectiveness of {\sc TDCleaner} with respect to other programming languages. 
We also plan to adapt {\sc TDCleaner} to other types of task comments, such as FIXME, HACK, etc.

\section*{ACKNOWLEDGMENT}
\label{sec:ack}
This research was partially supported by the ARC Laureate Fellowship FL190100035, and the National Research Foundation, Singapore under its Industry Alignment Fund – Prepositioning (IAF-PP) Funding Initiative. Any
opinions, findings, and conclusions, or recommendations expressed
in this material are those of the author(s) and do not reflect the
views of the National Research Foundation, Singapore. The authors would like to thank the reviewers for the insightful and constructive feedback.

\balance
\bibliographystyle{ACM-Reference-Format}
\bibliography{samples}

\end{document}